# Pomeranchuk instability in the nematic phase of two monolayer $FeSe/SrTiO_3$

C. Y. Tang[1,2,3*], X.-L. Peng[1,2*], Y.-H. Yuan[4], P. Zhang[5], G.-N. Phan[1], S.-Y. Gao[1,2], Y.-B. Huang[6], L.-Y. Kong[1], T. Qian[1,2], W. Li[4], Q.-K. Xue[4,8], Z.-Q. Wang[7], K. Jiang[1†], Y.-J. Sun[1,8†] & H. Ding[1,9]

[1]*Beijing National Laboratory for Condensed Matter Physics, and Institute of Physics, Chinese Academy of Sciences, Beijing 100190, China*

[2]*School of Physics, University of Chinese Academy of Sciences, Beijing 100190, China*

[3]*State Key Laboratory of Surface Physics and Department of Physics, Fudan University, Shanghai 200438, China.*

[4]*State Key Laboratory of Low-Dimensional Quantum Physics, Department of Physics, Tsinghua University, Beijing 100084, China*

[5]*School of Physics, Nanjing University, Beijing 210093, China*

[6]*Shanghai Advanced Research Institute, Chinese Academy of Sciences, Shanghai 201204, China*

[7]*Department of Physics, Boston College, Chestnut Hill, MA 02467, USA*

[8]*Department of Physics, Southern University of Science and Technology, Shenzhen 518055, China*

[9]*Tsung-Dao Lee Institute & School of Physics and Astronomy, Shanghai Jiao Tong University, Shanghai 200240, China*

[*] These authors contributed equally to this work

[†] Corresponding authors: jiangkun@iphy.ac.cn; sunyj@sustech.edu.cn

**Nematicity, where electrons break rotational symmetry while preserving translational symmetry, is ubiquitous in strongly correlated quantum matters, including high-$T_c$ cuprates and iron-based superconductors. A central question in nematicity is whether it is driven by Pomeranchuk instability in momentum space or orbital order (polarization) in real space, especially as nematicity intertwines with superconductivity. FeSe/$SrTiO_3$ (STO), where nematicity occurs without long-range magnetic order, is an ideal platform for studying the nature and origin of the electronic nematicity. Here we present direct evidence of Pomeranchuk nematic order in two monolayer FeSe/STO using angle-resolved photoemission spectroscopy, revealing a remarkable degeneracy of $d_{xz}$ and $d_{yz}$ bands at the Brillouin zone center, but a significant band separation at the zone corner. This momentum-dependent nematicity demonstrates that nematicity in FeSe/STO originates from the Pomeranchuk instability, offering insights into the relationship between nematicity and superconductivity. Our results establish two-dimensional FeSe thin film as a powerful platform for investigating quantum physics under complex intertwinement.**

Electronic nematicity, characterized by the breaking of lattice rotational symmetry while preserving translational symmetry in quantum electronic states, has garnered significant attention due to its prevalence in strongly correlated systems, including copper oxide superconductors [1,2], iron-based superconductors (FeSCs) [3-6], and twist bilayer graphene [7]. The ubiquitous presence of nematicity and its intimate relationship with the superconducting regime in phase diagrams suggest that understanding nematicity is critical to unraveling the microscopic mechanisms of unconventional superconductivity [6-11]. In this regard, FeSCs provide an ideal platform for studying electronic nematicity, due to their pronounced nematic phase transition, universality across diverse material families, and high tunability though doping and pressure. Among them, the FeSe/$SrTiO_3$ (STO) system has an unprecedented high superconducting critical temperature ($T_c$), and a nematic transition that occurs without the onset of long-range magnetic order, unambiguously representing the cleanest demonstration of electronic nematicity. While nematicity in one monolayer (ML) system is suppressed by high-temperature superconducting transition, superconductivity does not take place and cannot be even proximitized to 2 ML system, making 2 ML FeSe/STO an ideal platform for studying the nematic phase.

To understand the nematic order, the central question lies in addressing the driving mechanism of nematicity [12]. Phenomenologically, there are two types of driving mechanisms of the nematic order in FeSe: the Pomeranchuk instability-induced Fermi surface (FS) distortion in momentum space [9,13] and orbital order (polarization) in real space arising from the occupation imbalance of degenerate atomic orbitals [14,15]. For the Pomeranchuk instability [13], quasiparticle interactions in the angular momentum $l = 2$ channel cause the FS to deform, resulting in rotation symmetry breaking FSs invariant under inversion in momentum space, as shown in Fig. 1(a). On the other hand, it is also possible to generate nematicity from the real space perspective. Under $C_4$ rotation symmetry, the rotation-related $d_{xz}$ and $d_{yz}$ orbitals host equal occupation. When an onsite/local orbital-dependent energy difference lifts the degeneracy and causes an imbalance in orbital occupation, it leads to an orbital polarization favoring the $d_{yz}$ orbital, as illustrated in Fig. 1(b). The consequences of the above two mechanisms are different. For the orbital polarization, it is local and homogeneous in real space, resulting in an isotropic degeneracy splitting of the energy levels of $d_{xz}$ and $d_{yz}$ orbitals in momentum space. Hence, an isotropic band separation ($\Delta_\Gamma = \Delta_M \neq 0$) opens at both the Γ and M point, as shown in Fig. 1(d). For the Pomeranchuk nematic order, the symmetry breaking in the $l = 2$ angular momentum channel has lattice harmonics that are anisotropic in momentum space, being zero at the Γ point and maximum at the M point. Therefore, the Pomeranchuk nematicity only generates maximum bands splitting at the M with $\Delta_M \neq 0$ around which the FSs in FeSe/STO locates, as illustrated in Fig. 1(c), but leaves the bands at Γ degenerate with $\Delta_\Gamma = 0$. Note that in different domains, the orbital characters of $d_{xz}$ and $d_{yz}$ swap, altering the band separation energy, as indicated in Fig. 3(e). In this work, we isolate the critical role of the Pomenranchuk instability in the nematic phase of 2 ML FeSe/STO, and provide direct evidence that nematicity predominantly originates from the Pomeranchuk instability-induced FS distortion in momentum space. Our work establishes FeSe thin films as an ideal platform for exploring profound quantum physics, including intertwined order, nematicity, and superconductivity.

The nematic order emerges in FeSe thin films on STO substrates when the temperature decreases below the nematic transition temperature $T_{nem}$ [16,17], and the nematicity enhances with decreasing layer thickness. As shown in Fig. 1(e), stripes develop in the vicinity of impurities in the 20 ML FeSe film. Upon reducing the film thickness to 2 ML, the nematicity becomes more pronounced, exhibiting smectic-like features [18], as illustrated in Fig. 1(g). The direction of these stripes changes 90° across the domain walls, signifying the breaking of the four-fold rotational symmetry within each domain, providing direct evidence of electronic nematicity. Notably, stripes have not been observed in 1 ML FeSe/STO [19], likely due to the heavy electron doping and strong electron-phonon coupling from the substrate, driving the material into the high-$T_c$ superconducting state [20]. In contrast, the high-$T_c$ state does not persist or even proximitized to the second layer of FeSe, which is weakly coupled to the first layer and can be evaporated by low-temperature annealing, akin to the second layer of epitaxial grown graphene [21-24]. The Fermi surface of 2 ML FeSe, as shown in Fig. 1(f), is consistent with previous results [11,21]. It should be noted that the hole-like bands around Γ point do not cross the Fermi level. The finite spectroscopic intensity observed near Γ is not from FS but from measurement resolution broadening. Figure 1(h) illustrates the band structure measured along the Γ - M direction for 2 ML FeSe, with the hole-like bands at Γ situated below the Fermi energy ($E_F$), indicating that the FS of 2 ML FeSe solely around the zone corner. The nematicity induces a band separation at the M point between the two hole-like bands (see also Fig. S1 in Supplementary Materials), defined as $\Delta_M$, mirrors the energy separation observed in detwinned FeSe single crystals [25-27]. Notably, the band separation size is ~70 meV, surpassing the ~50 meV detected in bulk FeSe at 30 K [26].

In angle-resolved photoemission spectroscopy (ARPES) measurements, photons with different polarizations can excite electrons with distinct wavefunction symmetries within crystals [28]. The experimental geometry defines the *p*- and *s*-polarization, as shown in Fig. 2(a). Odd (even) orbital with respect to the $M_x$ mirror plane are selectively excited by *p*- (*s*-) polarized photons. The band dispersion along the Γ - M direction measured by *p*-polarized photons is shown in Fig. 2(b). Two hole-like bands α and β are clearly observed with different photon energies (*hv* = 26, 28, 32, and 42 eV, respectively). As the photon energy increases, the intensity of the α band decreases while that of the β band increases, attributed to the opposite photoemission cross sections of Fe 3*d* and Se 4*p* orbitals in the energy range of our experiments [29,30]. Therefore, we can infer that the β band has Fe $d_{yz}$ orbital character and the α band partially comprises of the Se $p_z$ orbital character. On the other hand, when the exciting photons are switched to *s*-polarization, only the α band is observed, as shown in Fig. 2(e), confirming its $d_{xz}$ orbital character.

Significantly, unlike in bulk FeSe, the energy separation between the α and β bands at the Γ point ($\Delta_\Gamma$) in 2 ML FeSe vanishes, as shown in Figs. 2(b) and 2(c). To elucidate the energy separation more accurately, we present the energy distribution curve (EDC) at the Γ point, using data collected with 32 eV *p*- and *s*-polarized photons, as illustrated in Fig. 2(d). We choose this exciting energy because both energy bands are enhanced, as shown in Fig. 2(b)(iii) and Fig. 2(e)(iii). It is evident that these two bands are degenerate. Furthermore, to quantitatively determine the band dispersion, we extract the peak position of the momentum distribution curves (MDCs) and fit the data points with a parabola function, as illustrated in Fig. 2(f). The fitting results reveal that the band tops of the α and β bands are located at 9.39 ± 0.31 meV and 9.83 ± 1.05 meV below the Fermi energy ($E_F$), respectively. This further proves the degenerate nature of the $d_{xz}$ and $d_{yz}$ bands at the Γ point. On general grounds, there are typically two common origins for the band separation $\Delta_\Gamma$ at Γ point, orbital polarization and spin-orbital coupling (SOC). However, these two mechanisms cannot cancel each other out since they are coupled to different scattering channels. Therefore, the absence of the nematic band separation of the $d_{xz}$ and $d_{yz}$ bands at Γ point rules out the orbital polarization scenario. On the other hand, the significant energy separation around M point manifests the anisotropic nature of nematicity in the momentum space in 2 ML FeSe, consistent with the Pomeranchuk instability induced FS distortion, given that the FSs only exist around M point.

To further investigate the evolution of nematic order from an ultrathin film to bulk FeSe, we measured the band structure of FeSe films with thickness of 4, 20 and 60 MLs. Figure 3 represents the evolution of $\Delta_\Gamma$ as a function of film thickness. As shown in Fig. 3(a), the degeneracy of the two hole-like bands is lifted in the 4 ML film, and the band separation increases systematically with film thickness. The energy difference $\Delta_\Gamma$, as shown in Fig. 3(b), gradually increases from 0 meV in the 2 ML film to 32 meV in the 60 ML FeSe film, a value comparable to that observed in bulk FeSe samples [25,31,32]. This indicates that nematic splitting emerges in multilayer films at the Γ point.

In Fig. 3(c), employing 32 eV *p*- and *s*-polarized photons, we observe two sets of hole-like bands near the Γ point in 60 ML FeSe. It should be noted that the sample contains twined nematic domains, hence the two sets of energy bands belong to the $d_{xz}$ and $d_{yz}$ orbitals of different twin domains defined by their crystal coordinates, as shown in Fig. 3e(ii). In contrast, we only observe one set of energy bands in the 2 ML film, due to the absence of nematic order at Γ, as illustrated in Fig. 3(e)(i). Therefore, this observation is consistent with the Pomeranchuk instability nature of the nematicity in the 2 ML film. The band separation ($\Delta_\Gamma$) of the 60 ML film measured by two different photon polarizations has the same value, as depicted in Fig. 3(d).

To elucidate the band evolution more clearly, we summarize the binding energies of the $d_{xz}$ and $d_{yz}$ bands measured by 22 eV and 32 eV photons at the Γ point in Fig. 4(a). In thicker films, the binding energy of each band varies when measured with different photon energies, indicative of band dispersion along $k_z$. Despite the $k_z$ dispersion, the nematicity-induced band separation remains constant, as depicted in Fig. 4(b), where the results of $\Delta_\Gamma$ and $\Delta_M$ are presented as a function of film thickness. With increasing thickness, both $\Delta_\Gamma$ and $\Delta_M$ shift to higher values nearly in parallel, suggesting the presence of orbital polarization ($\Delta_\Gamma = \Delta_M \neq 0$).

In light of our results, the degeneracy of the $d_{xz}/d_{yz}$ bands at Γ in 2 ML films indicates that the Pomeranchuk instability is the primary driving force of the nematic transition. Furthermore, thinner films exhibit larger $\Delta_M$, smaller $\Delta_\Gamma$, and higher nematic transition temperatures [9,31-34]. With increasing layer thickness, orbital polarization contributes significantly to the nematicity in FeSe. Accordingly, the sign-change momentum-dependent nematic order observed in bulk FeSe likely comprises a predominant Pomeranchuk nematic order and thickness-dependent orbital polarization. Although a previous proposal suggested that three orbital orders could lead to accidental degeneracy [35], this hypothesis can be ruled out by the Γ point degeneracy at high temperatures, since distinct orbital orders should exhibit different temperature dependencies (see Fig. S2 in Supplementary Materials). Regarding the microscopic origin of the Pomeranchuk instability observed in our study, one plausible mechanism is that quantum fluctuations induced by inter-atomic Coulomb repulsion leads to a renormalized band structure where the van Hove singularity at M sits close to the Fermi level, facilitating the emergence of the *d*-wave bond nematic order [9]. Our observation of the nematic order with the *d*-wave form ($\Delta_\Gamma$ = 0, $\Delta_M$ ≠ 0), is consistent with the theoretical predictions. Alternatively, the Pomeranchuk instability may be driven by spin fluctuation mediated Landau interaction in the $l$ = 2 charge sector [6,36,37], which needs to be further verified. Our results provide concrete constraints on the possible microscopic origin of the electronic nematicity in FeSe [6].

Intriguingly, the SOC gap is absent at the Γ point in 2 ML FeSe/STO, nor clear band hybridization between $d_{xy}$ and $d_{xz}/d_{yz}$ has been observed. It is further evidenced by the strongly sensitive intensity ratio of the *xz*- and *yz*-character to the switching of the light polarization, in sharp contrast to what is expected in the presence of orbital mixing due to SOC. The absence of the SOC gap prompts questions regarding the mechanism responsible for the lifting of the $d_{xz}/d_{yz}$ degeneracy at the Γ point with increasing layer thickness. In iron-based superconductors, SOC has been extensively studied and discussed, it can induce a gap at the Γ point without breaking rotational symmetry [38-41]. One natural possibility is that the SOC strength in the 2 ML film is significantly weakened compared to other iron-based materials. A previous ARPES study has indicated that the SOC strength is material dependent in iron-based superconductors [38]. Although the atomic SOC strength remains constant, the effective SOC strength in materials largely depends on crystal field, electron correlations [42] and possibly spin fluctuations [43-46], all of which may exhibit thickness-dependent behavior. The SOC effect in FeSe thin films warrants further theoretical investigation.

We achieved an unprecedented understanding of the nematicity in FeSe, unambiguously settled the debate on its origin by observing the *d*-wave Pomeranchuk instability in 2 ML FeSe/STO, which exhibits the strongest nematicity. The Pomeranchuk instability induces a nematic distortion of FSs and lifts the degeneracy between the $d_{xz}$ and $d_{yz}$ orbitals around the M point, as shown in Fig. 4(c). Conversely, the $d_{xz}$ and $d_{yz}$ orbitals below the Fermi level are degenerated at Γ point, ruling out the orbital order scenario, as illustrated in Fig. 4(c). Further investigation of thickness-dependent FeSe thin films reveals that the momentum-dependent nematic order observed in bulk FeSe can be attributed to two primary components: the predominant *d*-wave Pomeranchuk instability-induced FS distortion and the thickness-dependent orbital polarization.

Our findings establish a novel framework for probing the interplay between nematicity and high-$T_c$ superconductivity, imposing stringent constraints on candidate microscopic mechanisms of high-$T_c$ pairing. Beyond identifying pure Pomeranchuk nematicity, the 2 ML FeSe/STO system exhibits intriguing emergent phenomena: the FS undergoes a dramatic topological reconstruction in 2 ML films, while the high-$T_c$ state observed in 1 ML FeSe/STO vanishes entirely in 2 ML, concomitant with the onset of a smectic phase that amplifies nematic correlations. This sharp dichotomy between monolayer and bilayer behavior underscores the critical role of dimensionality in stabilizing competing quantum phases. Notably, recent work has identified nematic fluctuations mediated superconductivity in the $FeSe_{1-x}S_x$ system [10]. The isolated nematic phase can be suppressed by sulfur substitution at a quantum critical point, where nematic fluctuations are most pronounced. Introducing sulfur into our 2 ML FeSe/STO system presents a promising avenue to systematically tune nematic order, resolving whether superconductivity is enhanced, suppressed, or coexists with modified nematicity. Furthermore, many unconventional features of high-$T_c$ materials, including electronic nematicity, originate from their quasi-two-dimensional nature, where quantum fluctuations are markedly enhanced. Our findings highlight that a two-dimensional system not only provides fresh perspectives on the behavior of correlated quantum materials under complex intertwined orders, but also sheds light on fundamental physics in the two-dimensional limit.

## METHODS

High-quality FeSe films for *in situ* ARPES measurements are grown on 0.7 wt% Nb-doped $SrTiO_3$ (001) substrates after degassing for 10 hours at 600 °C and then annealing for 1.5 hours at 950 °C in an ultrahigh vacuum molecular beam epitaxy (MBE) chamber. Substrates are kept at 310 °C during film growth. Fe (99.98%) and Se (99.999%) are co-evaporated from Knudsen cells. The flux ratio of Fe to Se is 1:10, which is measured by a quart crystal balance. The growth rate determined by the Fe flux, is fixed at 0.7 UC/min. During the growth process, the sample quality is monitored using reflection high-energy electron diffraction (RHEED). After growth, the FeSe films are annealed at 370 °C for 10 hours. Then, the samples are transferred *in situ* to the ARPES chamber for measurements. The ARPES measurements are conducted at the BL-09U "Dreamline" beamline of the Shanghai Synchrotron Radiation Facility (SSRF) using a VG DA30 electron analyzer under an ultrahigh vacuum better than $5 \times 10^{-11}$ torr. The energy resolution is set to ~12 meV for the band structure and ~16 meV for FS mapping, while the angular resolution is set to 0.2°. The spectra are recorded at 30 K unless specified otherwise. The scanning tunneling microscopy (STM) measurements are performed in an ultrahigh vacuum STM (Unisoku) with a base temperature of 4.2 K. A polycrystalline PtIr STM tip is used and calibrated on Ag islands before the STM experiments.

**ACKNOWLEDGEMENTS**

We thank J.-P. Hu, X.-X. Wu, R. Yu, and H. Miao for useful discussions, as well as W.-H. Fan and Y. Liang for technical assistance. This work is supported by grants from the Ministry of Science and Technology of China (2016YFA0401000, 2016YFA0300600, 2016YFA0302400, 2016YFA0300600) and the National Natural Science Foundation of China (11888101,11622435, U1832202). Z.Q.W. is supported by the U.S. Department of Energy, Basic Energy Sciences Grant No. DE-FG02-99ER45747.

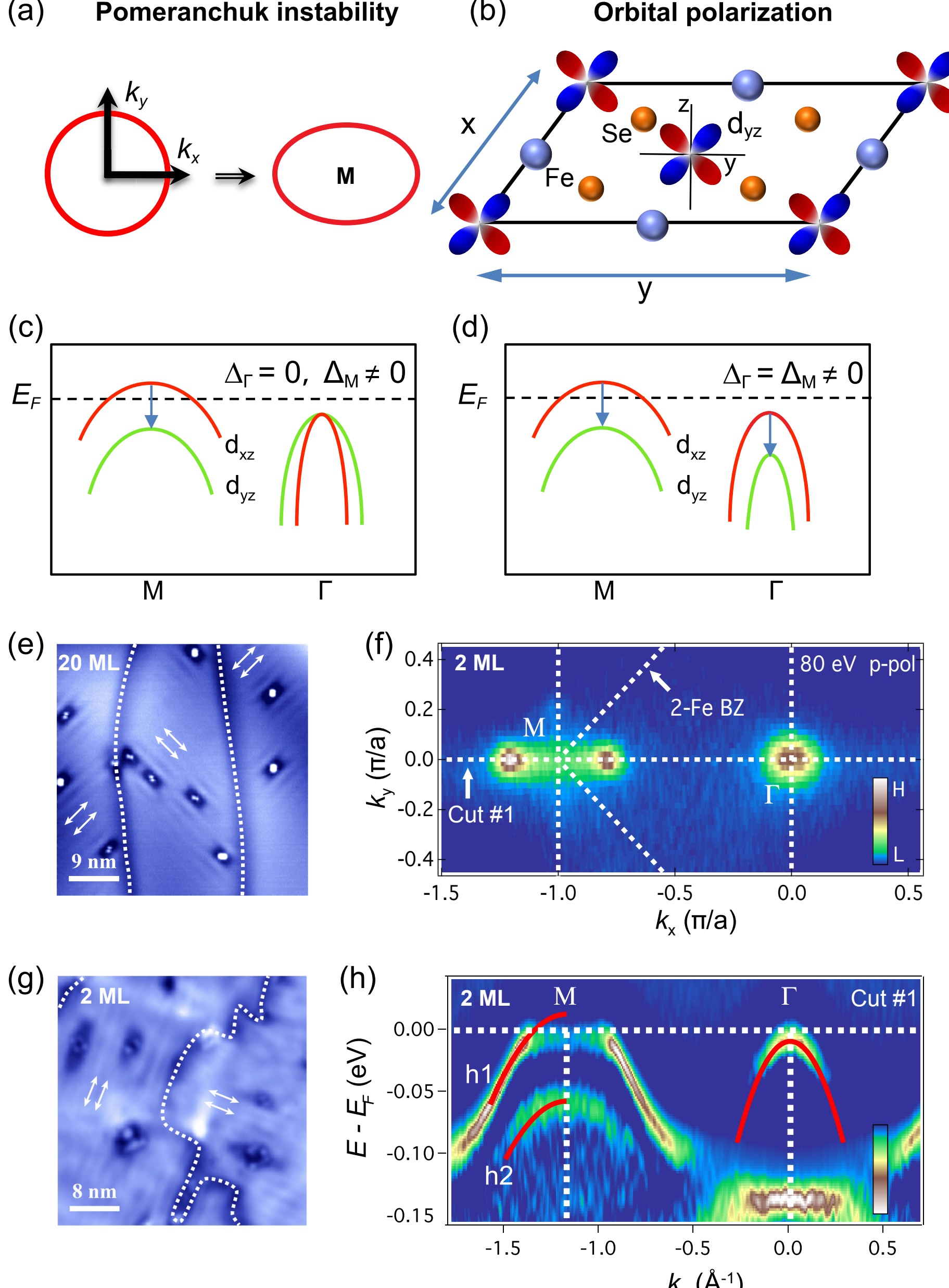

FIG. 1. Two types of nematicity in FeSe thin films on $SrTiO_3$ (STO). (a) Schematic of Pomeranchuk instability induced by Fermi surface distortion in the momentum space and (b) imbalance occupation in $d_{xz}/d_{yz}$ orbital induced orbital polarization in the real space, respectively. (c) Schematic of $d_{xz}/d_{yz}$ band degeneracy splitting at Γ and M driven by Pomeranchuk instability, and (d) orbital polarization, respectively. (e) STM topographic image of a 20 monolayer (ML) FeSe film on STO (45 nm × 45 nm; $V = 60$ mV, $I = 40$ pA). The white arrows indicate the stripes with twofold symmetry. The direction of stripes rotates 90° when crossing the domain walls (the white dashed lines). (f) The Fermi surface of 2 ML FeSe/STO

measured by 80 eV *p*-polarized photons at T = 20 K. The intensity has been integrated in the ±10 meV energy range. (g) same as e but for a 2 ML FeSe film on STO (40 nm × 40 nm; $V$ = 60 mV, $I$ = 100 pA). (h) Second-derivative spectrum of the band structure of 2 ML FeSe/STO measured by 80 eV *p*-polarized photons along the Γ - M direction. The red lines represent the energy band near the Γ and M points. Bands h1 and h2 denote two hole-like bands.

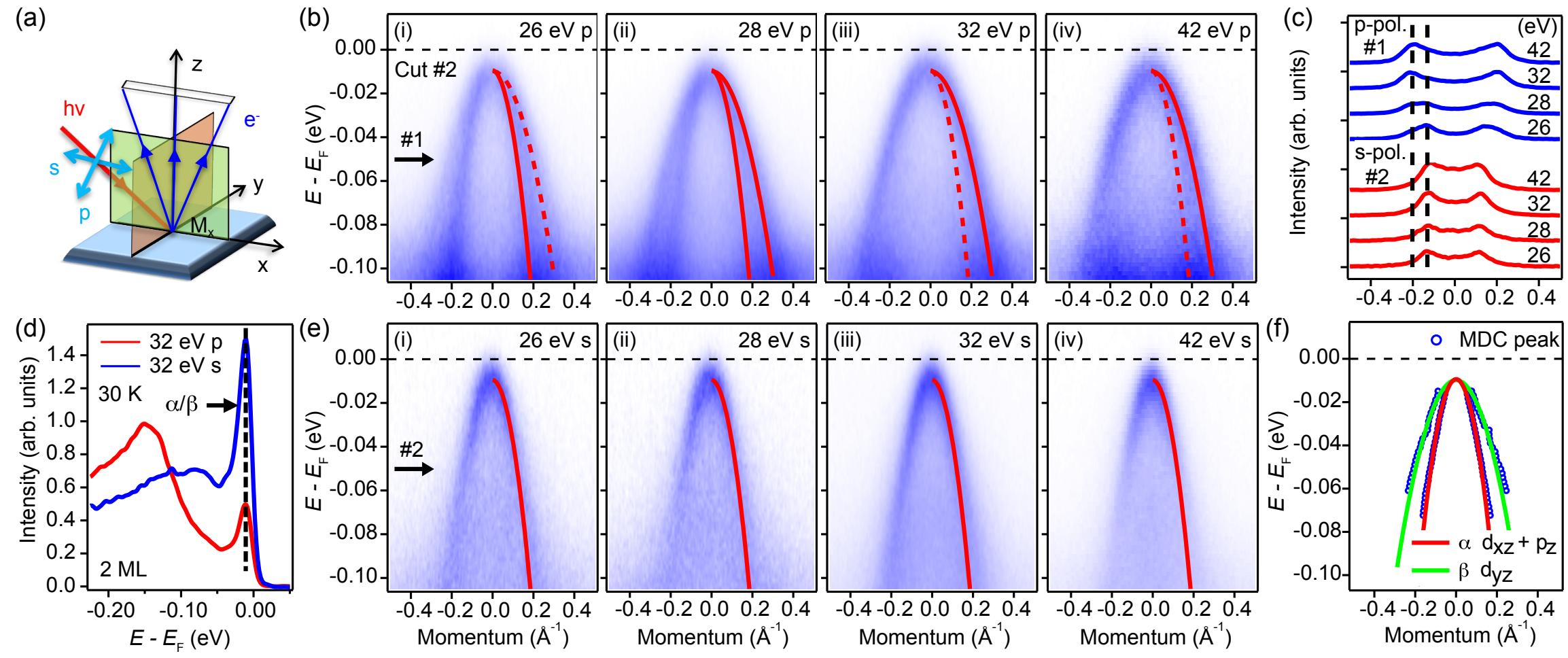

FIG. 2. Degeneracy of the $d_{xz}/d_{yz}$ orbitals at the Brillouin Zone center (Γ) in 2 ML FeSe on STO. (a) Schematic experimental geometry for the ARPES measurements. (b) Band structure measured by (i) 26 eV, (ii) 28 eV, (iii) 32 eV and (iv) 42 eV *p*-polarized photons along the Γ - M direction. The red lines represent the fitting results of two hole-like bands. Solid and dashed lines indicate the visible and invisible bands, respectively. (c) Comparison of the momentum distribution curve (MDC) measured by different photon energies and polarizations at the energies indicated by #1 in (b) and #2 in (e). The black dashed lines indicate the peak position. (d) Comparison of the energy distribution curve (EDC) at the Γ point measured by 32 eV *p*- and *s*-polarized photons. The black dashed lines indicate the positions of the α and β band tops. (e) Similar to (b) but measured by *s*-polarized photons. (f) Summary of the band structure and orbital analysis around Γ. Blue circles represent the peak position of MDC fitted by the Lorentz function. Two solid curves are the fitting results of MDC peaks.

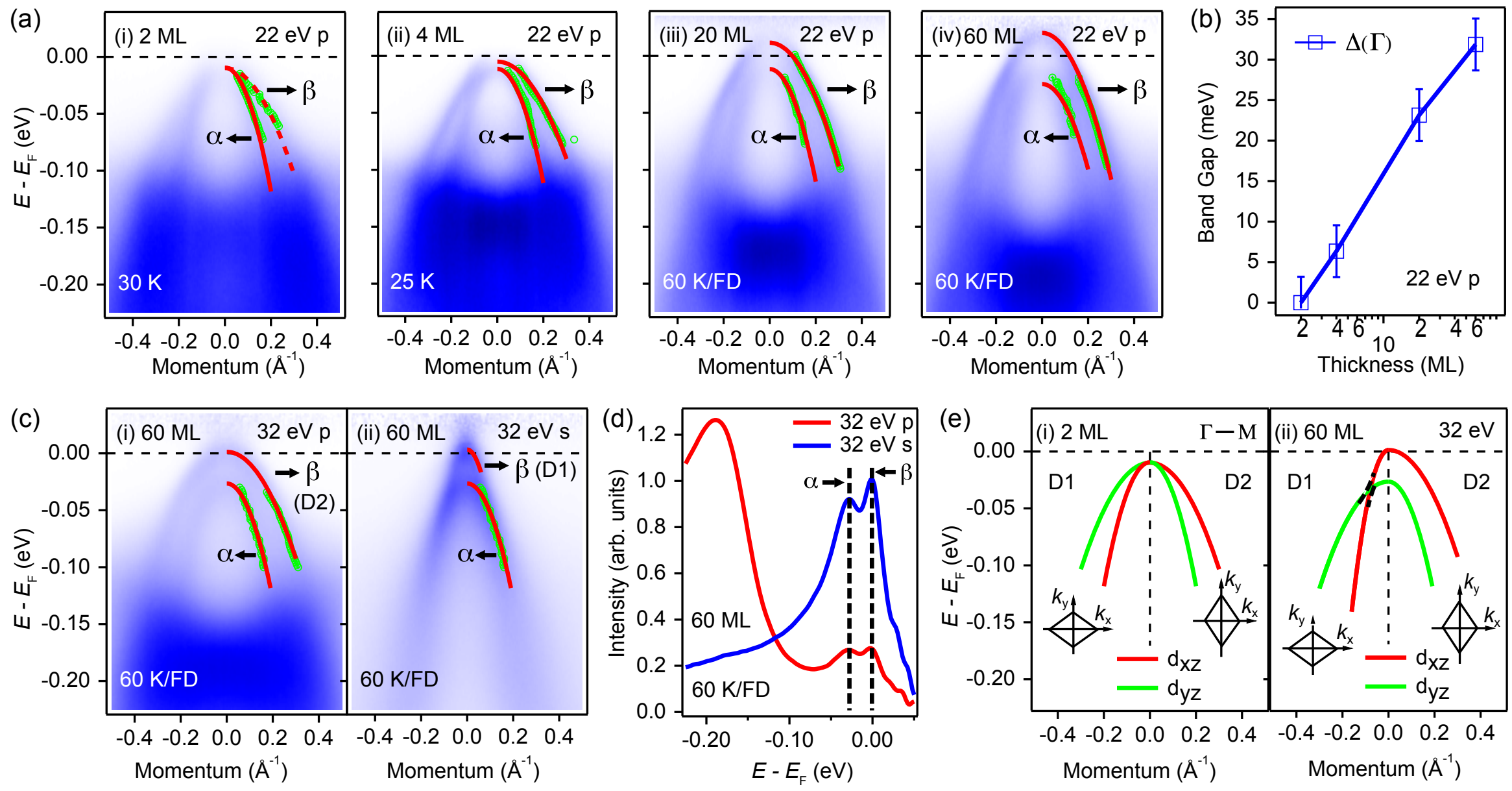

FIG. 3. Thickness-dependent $d_{xz}/d_{yz}$ band separation and nematic order around the Γ point. (a) Band structure of (i) 2 ML, (ii) 4 ML, (iii) 20 ML and (iv) 60 ML FeSe measured by 22 eV *p*-polarized photons along the Γ – M direction. The results of 20 ML and 60 ML are divided by the Fermi–Dirac distribution function convoluted by the resolution function to visualize the states above $E_F$. Red lines represent the fitting results of band dispersion. (b) Evolution of the band separation between α and β with film thickness. (c) Band structure along the Γ – M direction of 60 ML FeSe measured by 32 eV (i) *p*- and (ii) *s*- polarized photons. The red lines represent the band dispersion of the α and β bands. D1 and D2 represent domain 1 and domain 2. (d) Comparison of the energy distribution curve (EDC) at the Γ point of the data in (c). The black dashed lines indicate the positions of the α and β band tops. (e) Schematic band structure around the Γ point of (i) 2 ML and (ii) 60 ML for two perpendicular domains. The $k_x$-axis is defined along the 2-Fe Brillouin zone direction.

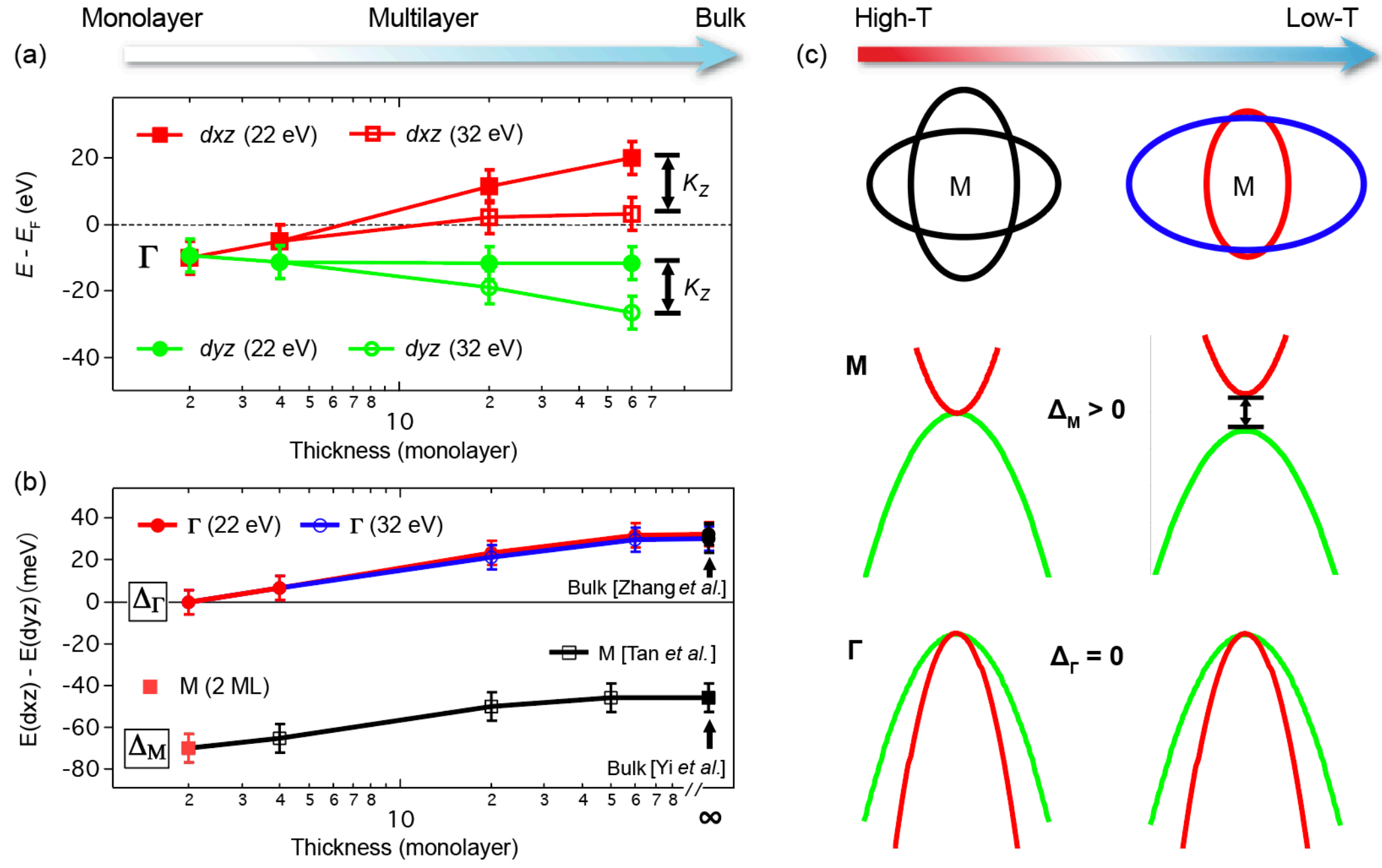

FIG. 4. Summary of thickness-dependent nematic order in FeSe films on STO. (a) Summary of the thickness-dependent binding energy of hole-like bands at the Γ point and the corresponding orbital character measured by 22 eV and 32 eV photons. $k_z$ represents the variation of the band top with photon energy. (b) Summary of the evolution of the onsite energy difference between the $d_{xz}$ and $d_{yz}$ orbitals at Γ and M point with film thickness. The data of bulk FeSe are adapted from ref. 31 for the Γ point and ref. 26 for the M point. The data of the M point of multilayer FeSe are adapted from ref. 11. (c) Schematic of the Pomeranchuk instability induced nematic distortion of FSs and the band structure near M and Γ point observed in 2 ML FeSe/STO, respectively.